# THERMAL ENTROPY BASED HESITANT FUZZY LINGUISTIC TERM SET ANALYSIS IN ENERGY EFFICIENT OPPORTUNISTIC CLUSTERING

Junaid Anees and Hao-Chun Zhang

School of Energy Science & Engineering,
Harbin Institute of Technology, Harbin, China

*ABSTRACT*

*Limited energy resources and sensor nodes' adaptability with the surrounding environment play a significant role in the sustainable Wireless Sensor Networks. This paper proposes a novel, dynamic, self-organizing opportunistic clustering using Hesitant Fuzzy Linguistic Term Analysis- based Multi-Criteria Decision Modeling methodology in order to overcome the CH decision making problems and network lifetime bottlenecks. The asynchronous sleep/awake cycle strategy could be exploited to make an opportunistic connection between sensor nodes using opportunistic connection random graph. Every node in the network observe the node gain degree, energy welfare, relative thermal entropy, link connectivity, expected optimal hop, link quality factor etc. to form the criteria for Hesitant Fuzzy Linguistic Term Set. It makes the node to evaluate its current state and make the decision about the required action ('CH', 'CM' or 'relay'). The simulation results reveal that our proposed scheme leads to an improvement in network lifetime, packet delivery ratio and overall energy consumption against existing benchmarks.*

## 1. INTRODUCTION

Wireless Sensor Networks (WSNs) should operate for a long time for a specific application like Smart Home, environmental monitoring, disaster management, forest fires, precision agriculture, surveillance system traffic monitoring etc. Power-constrained WSNs have to adjust their sleep/awake cycle according to the application requirements in order to maximize the network lifetime and overall energy consumption [1]. Sectional failure and thermal exposure can cause significant damage to sensor nodes. Moreover, different units of a sensor node behave differently when exposed in sunlight for long period of time for example; the performance of a typical transceiver is degraded with the increase in temperature. The purpose of deploying WSNs is to achieve a shared goal through sensor collaboration and data aggregation. In order to allocate the resources to sensor nodes effectively, topology architecture is needed in which sensors are organized in clusters [1]. The multi-hop routing in this clustering topology can result in the decrease of overall energy consumption and interference among sensor nodes due to specific timeslots allocation [2]. In addition to it, it could also effectively optimize the data redundancy by significantly reducing the collected data size using data aggregation techniques at Cluster Head (CH) level [1-2].

Researchers have proposed different node scheduling techniques to save battery power of sensor nodes i.e. synchronous and asynchronous sleep/awake scheduling. Asynchronous sleep/awake scheduling is designed to prolong the network lifetime and improve energy utilization by creating an opportunistic node connection between sensor nodes in the network [3-6]. A very popular technique to ensure sustainable operation of sensor nodes is Opportunistic Routing (OR) which is a paradigm in WSNs that benefit from broadcasting characteristics of a wireless medium by selecting multiple sensor nodes as candidate forwarders. This set of sensor nodes is called a Candidate Set (CS). The performance of OR significantly depends on several key factors, such as the OR metric, the candidate selection algorithm, and the candidate coordination method. Based on the asynchronous sleep/awake scheduling in OR, a node can sense, process, and transmit/receive during its active state [3-4]. Conversely, a node enters its sleeping state for an interval predefined or calculated according to contemporary environmental conditions. Researchers have also worked on temperature adaptive sleep/awake scheduling techniques [7-8].

The information entropy utilizes probability distribution function (pdf) to statistically measure the degree of uncertainty [9]. The entropy H(X) of a random variable $X = \{x1, x2, \ldots xn\}$ having probability distribution as p(X) can be given as $H(x) = \sum_{x \in X} p(x) log_2 p(x)$ for $0 \leq H(X) \leq 1$ [9-10]. It should be kept in focus that CH or BS should not be hesitant or irresolute about any of their decisions regarding cluster formation and data fusion. Keeping in view OR and temperature adaptive sleep/awake scheduling, we have selected multiple parameters including time frequency parameter, node's gain energy, relative thermal entropy, expected optimal hops, link quality factor in terms of signal-to-noise ratio, as our attributes of hesitant fuzzy linguistic term set. These attributes are used to assess the role of nodes and self adaptively make the appropriate decision in a round of operation. No concept of duty cycle is used in our proposed scheme. Furthermore, our proposed scheme FLOC uses this information in a Multi-Attribute Decision Modelling (MADM) framework to efficiently utilize our hesitant fuzzy linguistic term set to incorporate a qualitative assessment of the parameters by a node and help the node observe a situation adaptive role transition. The rest of the paper is organized as follows: Section 2 contains the discussion on some related works. System modelling is presented in Section 3. Our proposed scheme FLOC is presented in Section 4. HFLTS analysis is provided in Section 5. Section 6 presents the simulation framework and performance evaluation of the proposed technique. Finally, section 7 concludes the paper with some targeted future works.

## 2. RELATED WORK

Various researchers have focused on proposing different routing protocols for WSNs based on different parameters such as end–end delay, packet delivery ratio, network lifetime, overall energy consumption, control packet overhead, and sink node mobility, etc. Ogundile et al. [1] presented a detailed survey for energy efficient and energy balanced routing protocols for WSNs including the taxonomy of cluster-based routing protocols for WSNs. Routing protocols in WSNs can be segmented into two main categories, i.e., hierarchical and non-hierarchical routing protocols. Non-hierarchical routing protocols are designed in accordance with overhearing, flooding, and sink node position advertisement, whereas hierarchical routing protocols are designed on the basis of grid, tree, cluster and area [1-3][12]. Different hierarchical routing protocols have their own merits and demerits, but as far as cluster-based hierarchical routing protocols are concerned, researchers have been challenged with a task of achieving an optimal balance between end–end delay and energy consumption [9-13].

Yang et al. [5] introduced the utilization of sleep/awake cycle of sensor nodes to prolong the network lifetime. The sleep/awake cycle can be segmented into two categories—synchronous and asynchronous sleep/awake cycle. In this paper, our focus is only towards asynchronous sleep/awake scheduling. Depending on the network connectivity requirements in terms of traffic coverage, Mukherjee et al. [15] proposed an asynchronous sleep/awake scheduling technique with a minimum number of sensor nodes to achieve the required network coverage. As a result of asynchronous sleep/awake scheduling, opportunistic node connections are established between sensor nodes and their neighbours, which brings the need of Opportunistic Connection Random Graph (OCRG) theory to properly model the opportunistic node connections by forming a spanning tree. Anees et al. [6] proposed an energy-efficient multi-disjoint path opportunistic node connection routing protocol for smart grids (SGs) neighbourhood area networks (NAN). Anees et al. [13] also proposed a delay aware energy-efficient opportunistic node selection in restricted routing for delay sensitive applications. In this protocol, the information related to updated position of sink is advertised by multiple ring nodes and data is forwarded to mobile sink using ring nodes having maximum residual energy.

In a few asynchronous sleep/awake scheduling techniques, the sensor nodes are found to remain in active listening mode for a long amount of time, resulting in unnecessary consumption of energy. A popular WSN MAC protocol, Sensor Medium Access Control (SMAC), has been proposed by Ye et al. [16]. SMAC protocol lets the node listen for a fixed interval of time and turn their radio off (sleep state) for a fixed duration. Barkley-MAC (BMAC) [17] provides an adaptive preamble sampling technique to effectively reduce the duty cycle and idle listening by the sensor nodes. They are required to wake up periodically to check for ongoing communications. Shah et al. [18] devised a guaranteed lifetime protocol in which the sink node assigns sleep/awake periods for other nodes depending on residual energy, sleep duration, and coverage by the nodes. A mathematical model for temperature adaptive sleep/awake strategy is developed by Bachir et al. [8] with three proposed algorithms i.e. Stop Operate (SO), a Power control (PC), and Stop-Operate-Power-Control (SOPC). The sensor nodes running any of the algorithms are supposed to observe the contemporary state based on a pre-calculated relationship between node-density and temperature. Thermal entropy of the sensor nodes has been explored in the intelligent sleep-scheduling technique iSleep [19]. Reinforcement Learning based sleep-scheduling algorithm RL-Sleep has been proposed in [7] in which the authors have used a temperature model and Q-learning technique to switch the sleep/awake states adaptively, depending on the environmental situation.

It has been revealed through a detailed literature review that most of the clustering schemes consider energy efficiency, traffic distribution, or coverage-efficiency as the prime criteria for state-scheduling and decision modelling of sensor nodes instead of relative thermal entropy, temperature adaptability or hesitant fuzziness used for nodes' role transition etc. A few entropy based clustering schemes have been proposed in which entropy weight coefficient method is adopted for decision making in cluster-based hierarchical routing protocol [20-21].Multi-Criteria Decision Analysis (MCDA) and Multi-Attribute Sensors Decision Modelling (MADM) using entropy weight coefficients are also types of entropy weight-based multi-criteria decision routing [21]. Anees et al. [9] proposed hesitant fuzzy entropy based opportunistic clustering and data fusion algorithm for heterogeneous WSNs. In this algorithm, the local sensory data is gathered from sensor nodes by utilizing hesitant fuzzy entropy based multi-attribute decision modelling for cluster head election procedure. Zhai et al. [10] developed Hesitant Language Preference Relationships (HLPR) to improve the credibility of WSNs by fusing uncertain information and putting forward exact opinion about different WSN schemes.

Varshney [22] proposed an emerging concept of simultaneous wireless information and power transfer (SWIPT) in which both energy and data are transferred over RF links simultaneously. Guo et al. [23] utilized the concept the SWIPT to extend the network lifetime of a clustered WSN by wirelessly charging the relay nodes which are responsible to share data with BS. Zhou et al. [24] proposed dynamic power splitting (DPS) to adjust the power ratio of information encoding and energy harvesting in EHWSNs. Anees et al. [25] proposed harvested energy scavenging and transfer capabilities in opportunistic ring routing in which a distinguishing approach of hybrid (ring + cluster) topology is used in a virtual ring structure and then a two-tier routing topology is used in the virtual ring as an overlay by grouping nodes into clusters. Overall, to the best of our knowledge, there is no published literature which focuses on thermal entropy based HFLTS analysis for energy efficient opportunistic clustering. In this paper, we have considered a set of attributes that regulate the nodes' decisions about its role transition conducive to the current situation in a cluster and provided a detailed solution for optimally handling problems in energy efficient opportunistic clustering using relative thermal entropy based HFLTS analysis.

## 3. SYSTEM MODELING

### 3.1. Network model

A $MxM$ network area denoted as $A$ is considered for FLOC in which $N$ sensor nodes are deployed randomly and independently. We have assumed that sensor nodes follow a uniform distribution. The node-density of the network is denoted as $\lambda_0 = \frac{N}{A}$. All sensor nodes use short radio range (RS) for sensing and transmission purposes whereas sink node can use RS for transmission & reception and long radio range (RL) for data collection tasks using a tag message. However, all sensor nodes can exploit the power control function and communicate with different neighbouring nodes within various power levels. A probe message is shared by each sensor node to acquire the neighbour information as discussed in [6]. Each sensor node is equipped with a power splitting radio, which is composed of a signal processing unit to transfer energy to or from neighbours using RF link. Moreover, it is also assumed that every sensor node is aware of its position using the energy-efficient localization method [26-29].

Each sensor node is characterized by a set of k attributes named as $C = \{c_1, c_2, \ldots c_k\}$ and a set of weights $w_t = \{w_{t1}, w_{t2}, \ldots w_{tp}\}$ is assigned by sensor node to the $p$ criteria of $C$. Furthermore, the sensor node undergoes $y$ states i.e. $ST = \{ST_1, ST_2, \ldots ST_y\}$, where $ST_1$ represents the favorable state (attribute values are above threshold) and $ST_y$ represents the stressed state (attribute values below threshold. Depending on multiple parameters, the sensor node decides about the most suitable action against the contemporary state i.e. $AC = \{CH, CM, Relay\}$. Hesitant Fuzzy Sets have been used in our proposed scheme which enables the sensor node to decide about the optimal action after assessing the respective conditions.

### 3.2. Energy Model

The energy consumption model [6] for radio energy dissipation during transmission and reception is considered in which the energy required to transmit l bits of data over distance d can be given in (1) as:

$$E_{Tx}(V_i, V_j) = \begin{cases} E_{elec}l + \varepsilon_{fs}ld_{V_iV_j}^2 & d < d_0 \\ E_{elec}l + \varepsilon_{mp}ld_{V_iV_j}^4 & d \geq d_0 \end{cases} \quad (1)$$

where $E_{elec}$ is the energy spent by transmitter on running the radio electronics, $\varepsilon_{fs}$ is the free space energy dissipated by power amplifier depending on the Euclidean distance $d_{V_iV_j}$ between the transmitter and receiver, $\varepsilon_{mp}$ is the muti-path fading factor for energy dissipated by power amplifier depending on Euclidean distance $d_{V_iV_j}$ between transmitter and receiver. The threshold distance $d_o$ is given as $d_o = \sqrt{\varepsilon_{fs}/\varepsilon_{mp}}$. Likewise, the energy required to receive l bits of data over distance d is given in (2) as:

$$E_{Rx} = E_{elec}l \qquad (2)$$

The energy used for sensing l bits of data in the virtual ring at the beginning of each round can be given as $E_{sense} = E_{elec}l$. Accordingly, the total energy consumed by cluster member (CM) can be computed in (3) as:

$$E_{CM} = E_{sense} + E_{Tx} = E_{elec}l + E_{elec}l + \varepsilon_{fs}ld^2_{V_iV_j} \qquad (3)$$

Each CH is responsible for data gathering, aggregating the received data and then relaying that data towards sink, so the total energy consumed by a CH can be computed in (4) - (5) as

$$E_{CH} = E_{sense} + \left(\frac{N}{N_C} - 1\right)E_{Rx} + \left(\frac{N}{N_C}\right)lE_A + \left(\frac{N}{r}\right)E_{Tx} \qquad (4)$$

$$E_{CH} = E_{elec}l + \left(\frac{N}{N_C} - 1\right)E_{elec}l + \left(\frac{N}{N_C}\right)l\frac{E_{elec}}{R_{CC}} + \left(\frac{N}{r}\right)E_{elec}l + \left(\frac{N}{r}\right)\varepsilon_{mp}ld^4_{V_iV_j} \qquad (5)$$

where $N_C$ represents the number of clusters in the network, $\frac{N}{N_C}$ is the number of working sensor nodes per cluster in which we have 1 CH and $\frac{N}{N_C} - 1$ CMs. $E_A$ signifies the data aggregating energy at CH level, r represents the compression ratio and $R_{CC}$ symbolizes the communication to computation ratio.

## 4. PROPOSED SCHEME FLOC

### 4.1. Ambient temperature and Relative thermal entropy

In this section, the proposed scheme FLOC is discussed in detail. As the hesitant fuzzy linguistic term set analysis is based on MADM, we need to consider several parameters like ambient temperature, asynchronous sleep/awake cycle, relative thermal entropy, gain degree, expected optimal hops and link quality factor as attributes of hesitant fuzzy set. Keeping in view the diurnal temperature variation caused by solar radiation, the sensor nodes placed under direct sunlight absorb higher heat energy than the sensor nodes in shadow. According to temperature model in [7], the temperature of a sensor node $i$ after solar heat absorption for amount of time $\Delta t$ can be represented in (6) as,

$$T^i_{t+\Delta t} = max\left\{T^i_t + \frac{(S_{SUN}(t)\alpha(t) - \eta T^4)}{c_p\theta}Area_{sen}\Delta t, T^i_t\right\} \qquad (6)$$

where $T^i_t$ is the temperature of a node $i$ at time $t$, $S_{SUN}(t)$ denotes the amount of radiation by the sun at that time, $\alpha(t)$ is the temporal variation of sun exposure, $Area_{sen}$ is the exposed area through which the sensor node absorbs solar heat, $\eta$ is Boltzman constant, $\theta$ represents the mass of the sensor node, $c_p$ represents the specific heat and $T^i_{t+\Delta t}$ symbolizes the ambient temperature. The change in temperature of a sensor node can be extracted from equation (7) i.e.

$$\Delta T_i = \frac{(S_{SUN}(t)\alpha(t) - \eta T^4)}{c_p\theta}Area_{sen}\Delta t \qquad (7)$$

The resultant temperature $T_i$ of a sensor node i is given as $T_i = T_{t+\Delta t}^i = T + \Delta T_i$. Foregoing in view, the solar radiation pattern for a day can be represented as $S_{SUN}(t) = S_{SUN}^{max} exp^{\frac{-(t-\rho)^2}{2\sigma^2}}$, $0 \leq t \leq 2\rho$ where $S_{SUN}^{max}$ is the peak value of the solar radiation during the day [7]. It has been found that the identical sensor nodes behave differently to the temperature variations due to solar radiation exposure, traffic flow and relative position of the sensor node [8]. Let $T_i$ be the temperature of the $i$th node and $T_H$ be the highest temperature for which the $i$th node becomes non-operational. The probability of failure of a sensor node due to temperature increase can be represented as $p_i = \frac{T_i}{T_H}$, where $T_i$ can be acquired from equation (6) & (7) and $T_H$ symbolizes the highest temperature the sensor node can withstand. Here we have assumed that $T_H$ is the same for all sensor nodes in the network. The cumulative effect of failure likelihood leads to network instability; therefore we need a probability distribution function (PDF) to measure the degree of uncertainty in the sensor network. It should be kept in focus that any sensor node due to failure likelihood should not be hesitant or irresolute about any of the operating mode. This hesitancy or irresolution resembles entropy.

The entropy H(X) of a random variable $X = \{x_1, x_2, \dots x_n\}$ having probability distribution as p(X) can be given as $H(X) = -\sum_{x \in X} p(x) log_2 p(x)$ for $0 \leq H(X) \leq 1$ [9]. Similarly, the Shannon's entropy at $i$th node can be defined as $H(p_i) = -p_i log_2 p_i$. The relative contribution of a sensor node towards the probable instability of the network can be estimated using relative thermal entropy by calculating the entropy in neighborhood i.e.

$$H_{rel}^{therm} = \frac{H(p_i)}{\sum_{j \in nbr_i} H(p_j)} \qquad (8)$$

Where $H_{rel}^{therm}$ indicates the relative thermal entropy and $nbr_i$ represents the neighborhood dataset.

### 4.2. Energy transfer and Asynchronous sleep/awake cycle

The amount of energy a node $i$ could acquire from its neighbouring sensor node $j$ through RF transfer based on sensor node's ability to control their power level, can be defined in equation (9) and (10) as:

$$E_{trans(V_j,V_i)} = \eta_1 \mu P_j |h_{V_i,V_j}|^2 = \eta_1 \mu P_j |\beta_1 d_{(V_i,V_j)}^{-\alpha_1}|^2 \qquad (9)$$

$$\Gamma_{V_i} = \sum_{j=1}^{k} E_{trans(V_j,V_i)} = \sum_{j=1}^{k} \eta_1 \mu P_j |\beta_1 d_{(V_i,V_j)}^{-\alpha_1}|^2 \qquad (10)$$

where $E_{trans(V_j,V_i)}$ is the amount of energy node $j$ can transfer to its neighbor $i$, $\eta_1$ is the energy conversion efficiency $0 < \eta_1 < 1$, $\mu$ is the energy and data splitting ratio $0 < \mu < 1$, $P_j$ is the signal power received from node j, $h_{V_i,V_j}$ is the channel gain, $\beta_1$ is a constant which depends on the environment's radio propagation properties, $\alpha_1$ is the path loss exponent, and $\Gamma_{V_i}$ is the node $V_i$'s gain degree. Foregoing in view, the total available energy at node $i$ can be computed as:

$$E_{T(V_i)} = \Gamma_{V_i} + E_{Bat(V_i)} \qquad (11)$$

where $E_{Bat(V_i)}$ is the remaining battery energy of node i in (11). The amount of energy shared by a node with its neighbours depends on activities such as sensing, relaying, sleep/awake schedule etc. In contrast to conventional routing algorithms in WSNs, our proposed scheme can serve both data and energy in its routing topology. We used opportunistic connection random graph (OCRG) in FLOC to model the opportunistic node connections between sensor nodes. Let $G(S_{SN}, O_C, L)$ be the graph representing OCRG in which $S_{SN}$ represents the set of nodes in the network, $O_C$ represents set of opportunistic connections existing between any two adjacent neighbours and $L$ represents the link connectivity of any two adjacent nodes in $S_{SN}$. As we know that if any sensor node works for longer time, it is highly likely that sensor node will be able to communicate with neighbours due to higher status transition frequencies, thus it will contribute in improving the link connectivity. The link connectivity also depends on the data routing cost $DR_C: O_C \to R$ such that $DR_C(i,j)$ is the cost associated with link $(i,j)$. Our routing metric can be defined as: $Min \sum_{i=1}^{n-1} DR_{C(V_i, V_{i+1})}$, $DR_{C(V_i, V_{i+1})} \geq 0$. The data routing cost can be computed using $E_{C(V_i, V_j)}$, $E_{T(V_i)}$ and $E_{T(V_j)}$ and is given in (12).

$$DR_{C(V_i, V_j)} = \frac{E_{C(V_i, V_j)}}{(E_{T(V_i)} + E_{T(V_j)})} \qquad (12)$$

where $E_{C(V_i, V_j)}$ is the transmission energy consumed over link $(i,j)$, $E_{T(V_i)}$ is the available total energy (including battery and gained energy through RF transfer) of node $i$, $E_{T(V_j)}$ is the available energy (including battery and gained energy through RF transfer) of node $j$. It is pertinent to mention that energy is also transferred along with the data in the routing process to compensate for the transmission energy consumed over each link and more energy is conserved than consumed as the sensor nodes are using strong signals for transmission purposes.

As far as asynchronous sleep/awake cycle is concerned, we have proposed the concept of sleep/awake cycle schedule $(W_v/S_v)$ and status transition frequencies $(F_{ST})$ to investigate the opportunistic node connection between sensor nodes in each data collection period. We calculated the time-frequency parameter $TF_{V_i V_j}$ based on working time $W_{V_i}$ and $W_{V_j}$ of sensor nodes $V_i$ and $V_j$, data collection duration $T_{CP}$, status transition frequencies $F_{ST_i}$ and $F_{ST_j}$ in equation (13) and (14) as,

$$TF_{V_i V_j} = \left(\frac{F_{STV_i}}{F_{ST_{max}}} \times \frac{W_{V_i}}{T_{CP}}\right)\left(\frac{F_{STV_j}}{F_{ST_{max}}} \times \frac{W_{V_j}}{T_{CP}}\right) \qquad (13)$$

$$TF_{V_i V_{SINK}} = \left(\frac{F_{STV_i}}{F_{ST_{max}}} \times \frac{W_{V_i}}{T_{CP}}\right)(W_{V_{SINK}}) \qquad (14)$$

Using the time-frequency parameter $TF_{V_i V_j}$ and data routing cost $DR_C$, our link connectivity $L_{V_i V_j}$ can be computed in (15) as,

$$L_{V_i V_j} = \alpha_2 DR_C(i,j) + (1 - \alpha_2) TF_{V_i V_j} \qquad (15)$$

$$L_{V_i V_j} = \alpha_2 \left(\frac{E_{C(V_i, V_j)}}{(E_{T(V_i)} + E_{T(V_j)})}\right) + (1 - \alpha_2)\left(\frac{F_{STV_i}}{F_{ST_{max}}} \times \frac{W_{V_i}}{T_{CP}}\right)\left(\frac{F_{STV_j}}{F_{ST_{max}}} \times \frac{W_{V_j}}{T_{CP}}\right) \qquad (16)$$

Where $\alpha_2$ is the appropriate weight assigned to data routing cost and time-frequency parameter in (16).

## 5. HESITANT FUZZY LINGUISTIC TERM SET (HFLTS) ANALYSIS

We know that the nature of our problem is subjective and uncertain, that's why we need a fuzzy computation based technique like HFLTS. A generalization of the basic fuzzy set which deals with the uncertainty starting from the hesitation in the assignment of membership degrees of an element is known as Hesitant Fuzzy Set (HFS) [7-9]. We start the HFLTS analysis with a set of inputs containing total number of nodes, sink, neighbor information, context free grammar, transformation function, set of alternatives, set of criteria and weight assignment. In FLOC, a node can attain two states based on the node's gain energy and energy welfare. If the node's gain degree is greater than the threshold and the normalized energy welfare is greater than the half of the maximum value of energy welfare, then the state evaluation of a node can be considered as 'optimistic'. Likewise, if the node's gain degree is less than the threshold and the normalized value of energy welfare is less than the half of the maximum value of energy welfare, then the state evaluation of a node can be considered as 'pessimistic'. After evaluating the state and acquiring the neighbourhood information, the node calculates the relative thermal entropy with reference to neighborhood. The next step is to store different attributes like ambient temperature, relative thermal entropy, node gain degree, link connectivity, EOH [6] and link quality factor in an array and perform the data standardization by normalizing different attributes to obtain the fractional representation of attributes within [0 1] before defining the criteria. The set of required actions of a sensor node is known as alternatives which can be denoted as {'CH', 'CM', 'Relay'} and the suitable action chosen by the sensor node $i$ from alternatives is based on the criteria defined in (17) i.e.,

$$Criteria = \begin{Bmatrix} Node\ gain\ degree \\ Energy\ Welfare \\ Relative\ thermal\ entropy \\ Link\ Conn \\ Expected\ Optimal\ Hop \\ Link\ quality\ factor \end{Bmatrix} = \{E_{T(V_i)}, EW, H_{rel}^{therm}, L_{V_iV_j}, EOH, LQR\} \quad (17)$$

And the corresponding weights assigned to the members defining the criteria will be $w_T = \{w_1, w_2, \ldots w_{|Criteria|}\}$, $|Criteria|$ is the cardinality of Criteria. Now we assume our linguistic term set as,

$$S = \begin{Bmatrix} s_1:Extremely\ low(el) \\ s_2:very\ low(vl) \\ s_3:low(l) \\ s_4:medium(m) \\ s_5:high(h) \\ s_6:very\ high(vh) \\ s_7:perfect(p) \end{Bmatrix} \quad (18)$$

The normalized attribute values after data standardization in the hesitant fuzzy set are converted to linguistic term set S using triangular membership function as depicted from Fig 1. A context-free grammar $G_{CF}$ [7] has been used to produce linguistic terms for the alternatives against different values of the criteria. These HFS membership values are then transformed into HFLTS using a transformation function $E_{GCF}$ as shown in equation (19) and (20).

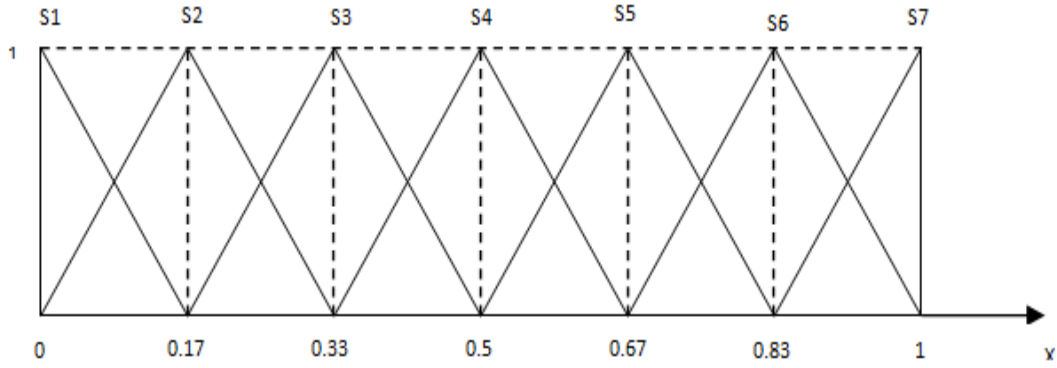

Figure1. Linguistic term set conversion using triangular membership function

$$H = \begin{array}{c} \\ x_1 \\ x_2 \\ x_3 \end{array} \begin{bmatrix} c_1 & c_2 & c_3 & c_4 & c_5 & c_6 \\ \text{greater than } h & \text{greater than } h & \text{lower than } m & \text{greater than } h & \text{lower than } m & \text{between } h \text{ and } p \\ \text{between } l \text{ and } h & \text{between } v_l \& h & \text{between } l \text{ and } v_h & \text{greater than } m & \text{between } l \text{ and } v_h & \text{between } m \text{ and } v_h \\ \text{greater than } l & \text{between } v_l \& h & \text{lower than } h & \text{greater than } m & \text{lower than } m & \text{between } h \text{ and } p \end{bmatrix}$$
(19)

where, $c_1$ = Node gain degree, $c_2$ = Energy Welfare, $c_3$ = Relative thermal entropy, $c_4$ = Link Connectivity, $c_5$ = Expected Optimal Hop, $c_6$ = Link quality factor, $x1$ = CH state, $x2$ = CM state, $x3$ = Relay state. Subsequently, the decision matrix $D_M$ is then converted to HFLTS by using a transformation function $E_{GCF}$.

$$H = \begin{array}{c} x_1 \\ x_2 \\ x_3 \end{array} \begin{bmatrix} c_1 & c_2 & c_3 & c_4 & c_5 & c_6 \\ \{v_h, p\} & \{v_h, p\} & \{el, v_l, l\} & \{v_h, p\} & \{el, v_l, l\} & \{h, v_h, p\} \\ \{l, m, h\} & \{v_l, l, m\ h\} & \{l, m, h, v_h\} & \{h, v_h, p\} & \{l, m, h, v_h\} & \{m, h, v_h\} \\ \{m, h, v_h, p\} & \{v_l, l, m\ h\} & \{el, v_l, l, m\} & \{h, v_h, p\} & \{el, v_l, l\} & \{h, v_h, p\} \end{bmatrix}$$ (20)

The decision matrix $D_M$ includes the members $h_{ij}$ where $i \in \{x_1, x_2, x_3\}$ and $j \in \{c_1, c_2, c_3, c_4, c_5, c_6\}$. According to the definition of hesitant fuzzy linguistic term set, we can easily calculate the envelope of its members $h_{ij}$ using upper bound and lower bound rules. Accordingly, the new decision matrix Y containing the envelopes of H is given in equation (21) as,

$$Y = \begin{array}{c} x_1 \\ x_2 \\ x_3 \end{array} \begin{bmatrix} c_1 & c_2 & c_3 & c_4 & c_5 & c_6 \\ \{v_h, p\} & \{v_h, p\} & \{el, l\} & \{v_h, p\} & \{el, l\} & \{h, p\} \\ \{l, h\} & \{v_l, h\} & \{l, v_h\} & \{h, p\} & \{l, v_h\} & \{m, v_h\} \\ \{m, p\} & \{v_l, h\} & \{el, m\} & \{h, p\} & \{el, l\} & \{h, p\} \end{bmatrix}$$ (21)

Now we utilize the node gain degree and energy welfare to classify the status of a node as 'Optimistic' and 'Pessimistic'. The 'RetainFunc' will be called if the status of a node is evaluated as 'Optimistic' and 'ChangeFunc' will be called if the status of a node is evaluated as 'Pessimistic' depending upon the status evaluation criteria already discussed.

| Function: RetainFunc (***Alternatives***, ***Criteria***, ***Y***) |
|---|
| 1. **For** $i = 1$ *to* $\|Alternatives\|$ of $Y$, |
| 2. **For** $j = 1$ *to* $\|Criteria\|$ of $Y$ |
| 3. Get 1-cut hesitant fuzzy set $H_S^j(xi)_{\alpha=1}$ for $H_S^j(xi)$, |
| where, $H_S^j(xi)_{\alpha=1} = [\{H_{S-}^j(xi)_{\alpha=1}, H_{S+}^j(xi)_{\alpha=1}\}]$ |
| 4. ***End*** |
| 5. ***End*** |
| 6. **For** $e = 1$ *to* $\|Alternatives\|$ of $Y$ |
| 7. **For** $f = 1$ *to* $\|Criteria\|$ of $Y$ |
| 8. ***Get*** the intervals $I_{max}(xe)$ *for each alternative* $x_i$ *with respect to each criterion* $f$; |
| where, $I_{max}(xe) = [Max(H_{S-}^f(xi)_{\alpha=1}), Max(H_{S+}^f(xi)_{\alpha=1})] f \in Criteria = [u_{e1}^{max}, u_{e2}^{max}]$ |
| 9. $Rank_{e1}^{opti} = \max(1 - \max(\frac{1-u_{i1}^{max}}{u_{i2}^{max}-u_{i1}^{max}+1}, 0), 0)$ (From equation (26)) |
| 10. ***End*** |
| 11. ***End*** |
| 12. ***Return*** $\max(Rank_{e1}^{opti})$ |

| Function: ChangeFunc (***Alternatives***,***Criteria***, ***Y***) ***Algorithm***: |
|---|
| 1. **For** $i = 1$ *to* $\|Alternatives\|$ of $Y$ |
| 2. **For** $j = 1$ *to* $\|Criteria\|$ of $Y$ |
| 3. Get 1-cut hesitant fuzzy set $H_S^j(xi)\alpha=1$ for $H_S^j(xi)$, |
| where, $H_S^j(xi)\alpha=1 = [\{H_{S-}^j(xi)\alpha=1, H_{S+}^j(xi)\alpha=1\}]$ |
| 4. ***End*** |
| 5. ***End*** |
| 6. **For** $e = 1$ *to* $\|Alternatives\|$ of $Y$ |
| 7. **For** $f = 1$ *to* $\|Criteria\|$ of $Y$ |
| 8. Get the intervals $Imin(xe)$ for each alternative $xi$ for each criterion f; |
| where $Imin(xe) = [Max(H_{S-}^f(xi)\alpha=1), Max(H_{S+}^f(xi)\alpha=1)] f \in Criteria = [u_{e1}^{max}, u_{e2}^{max}]$ |
| 9. $Rank_{e1}^{pessi} = \max(1 - \max(\frac{1-u_{i1}^{max}}{u_{i2}^{max}-u_{i1}^{max}+1}, 0), 0)$ (From equation (24)) |
| 10. ***End*** |
| 11. ***End*** |
| 12. ***Return*** $\max(Rank_{e1}^{pessi})$ |

The 'RetainFunc' and 'ChangeFunc' functions applies the 1-cut HFLTS to fuzzy sets in $Y$ to generate the envelope for each criteria against every alternative and calculates the probabilistic ranking of the alternatives based on the interval calculated from the envelopes. For instance, if the probabilistic ranking of alternatives is $[x_1 > x_3 > x_2]$, it indicates that the corresponding sensor node in its current state will probably retain its state and perform the role of a CH instead of CM or Relay node. But if the probabilistic ranking of alternatives is $[x_2 > x_3 > x_1]$ or $[x_3 > x_2 > x_1]$, it indicates that the sensor node's preferred action will be to change its role as CM or relay instead of CH.

## 6. RESULTS

### 6.1. Simulation Environment

We have evaluated the performance of FLOC in MATLAB 2019b and OMNET++ using cross platform library (MEX-API). This Application Programming Interface (API) can provide the user an easy bidirectional connection interface between MATLAB and OMNET++. Nodes are arranged in random topology. We have utilized low rate, low cost, short range, flexible and low power consumption standard IEEE 802.15.4 for our PHY and MAC layer. The performance metrics like active node ratio, average energy consumption and packet delivery ratio are analyzed against parametric benchmarks viz. node density and temperature variation. The performance of FLOC is compared with three different approaches i.e. 1) SOPC [8], 2) BMAC [17], 3) RL-Sleep [7]. Stop-Operate Power-Control (SOPC) is a temperature-aware asynchronous sleep-scheduling algorithm in which energy, link connectivity and network coverage are preserved by putting a few sensor nodes in hibernation mode and controlling the rest of the sensor nodes' transmission power. The communication range and number of active nodes are adjusted to maintain the critical density for consistent connectivity in the network. Berkley-MAC (BMAC) is a low-traffic, low-power-consuming MAC protocol based on adaptive preamble sampling for duty cycling to preserve energy, provide effective collision avoidance and high channel utilization. RL-Sleep is an asynchronous reinforcement learning based procedure based on the adaptive state transition determined by sensor nodes. The state transition is based on temperature sensing and collecting information from the neighbourhood. The effect of various parameters on the performance of FLOC with other existing benchmarks is provided in this section.

Table1. Simulation parameters

| Parameter | Value |
|---|---|
| Deployment area | 500 m X 500 m |
| $N$ | 60-90 |
| $T_H$ | 80°C |
| $S_{SUN}^{max}$ | 1 |
| Maximum Temperature | 80°C |
| $Area_{sen}$ | 20cm$^2$ |
| $d_0$ | 20m |
| $R$ | 200m |
| $\varepsilon_{fs}$ | 50nJ/bit/m$^2$ |
| $\varepsilon_{mp}$ | 10pJ/bit/m$^2$ |
| $E_{elec}$ | 50nJ/bit |
| Initial Energy for nodes | 5J(for neighbors of sink node) 3J (For other nodes) |
| $n_p$ | 2 |
| $c_p$ | 0.5 |
| $mass$ | 50g |
| $r$ | 0.25 |
| Number of packets | 1024 |
| Length of packet | 8000bits |

### 6.1.1. Active node ratio

Figure (2) depicts the active node ratio comparison of FLOC with SOPC, BMAC and RL-Sleep. It is evident from the figure that the ratio of average number of active nodes to total number of sensor nodes in the network is higher for FLOC than in any other benchmark. Furthermore, the active node ratio for all approaches is optimum for N=80. We also evaluated the performance of FLOC against SOPC, BMAC and RL-Sleep for varying diurnal temperature. Figure (3) shows the comparison of active node ratio of FLOC and other benchmarks for diurnal temperature variations. It has been observed that FLOC outperforms all three approaches in terms of active node ratio. The number of active sensor nodes in the network varies inversely with the diurnal temperature.

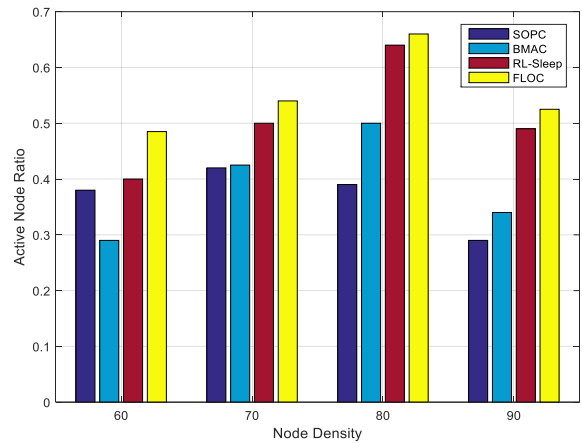

Figure 2. Active node ratio against node density

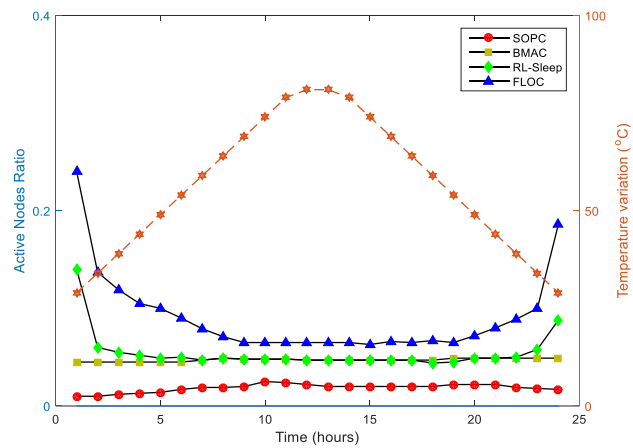

Figure 3. Active node ratio for diurnal temperature variations

### 6.1.2. Average energy consumption

Figure (4) shows the performance comparison of FLOC with SOPC, BMAC and RL-Sleep in terms of average energy consumption. BMAC outperforms all other algorithms due to its adaptive preamble strategy and short duty cycle which play a significant role in preserving energy. The adaptive adjustment of temperature with respect to communication range leverages higher energy consumption for SOPC. FLOC performs better than SOPC and RL-Sleep but exhibits higher amount of energy consumption against BMAC due to packet broadcasting in the neighborhood. Figure (5) depicts the average energy consumption of FLOC against other approaches for diurnal temperature variation. FLOC and BMAC exhibit almost similar profile for average energy consumed whereas SOPC and RL-Sleep consumed higher amount of energy for N=80.

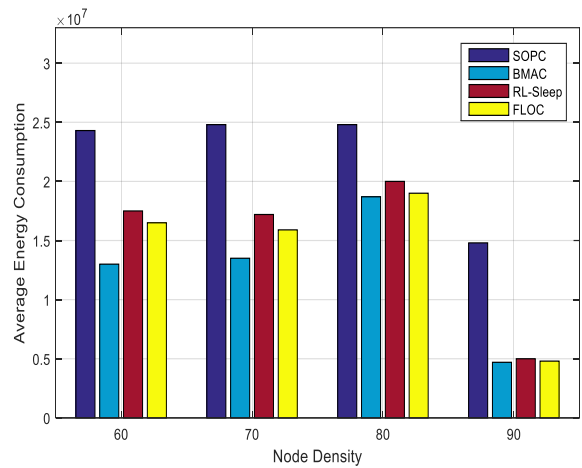

Figure 4. Average energy consumption against node density

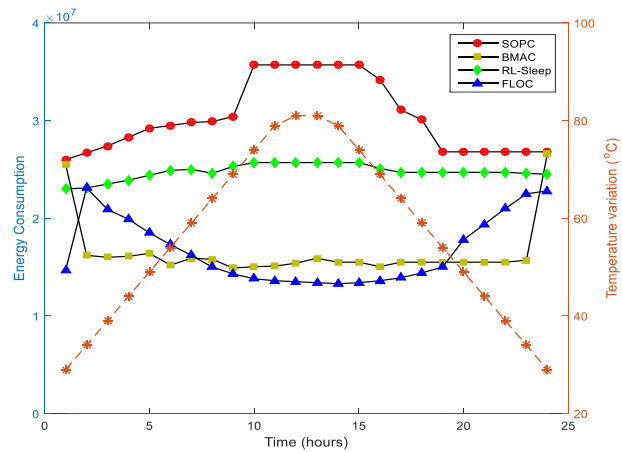

Figure 5. Average energy consumption for diurnal temperature variations

### 6.1.3. Packet Delivery Ratio (PDR)

Figure (6) depicts the comparison of FLOC with existing benchmarks in terms of PDR. FLOC outperforms other approaches in case of PD. Due to its opportunistic and environment adaptive sleep scheduling strategy, the additional power loss in FLOC can be compensated due to control packet overhead. BMAC shows the worst performance against existing benchmarks. Figure (7) shows the PDR of FLOC with other approaches for diurnal temperature variations. FLOC leverages a better packet delivery ratio in comparison to other techniques. It is pertinent to mention that PDR of FLOC decreases with the increase in diurnal temperature.

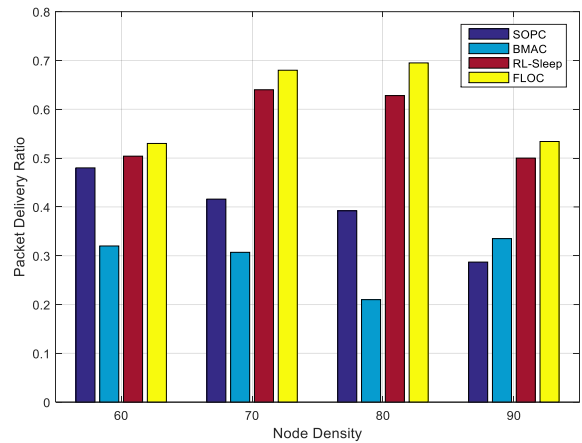

Figure 6. Packet delivery ratio against node density

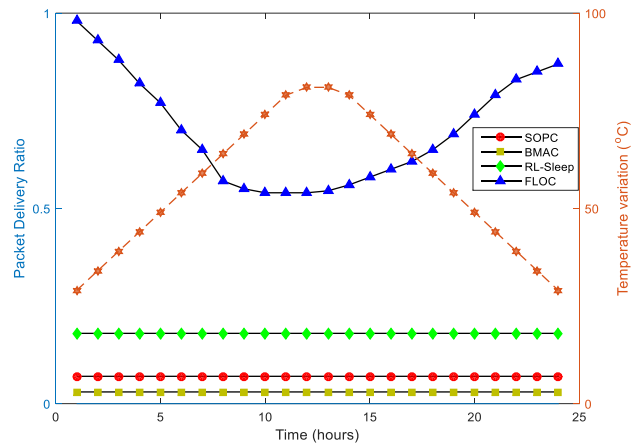

Figure 7. Packet delivery ratio for diurnal temperature variations

## 7. CONCLUSIONS

In this paper, a novel, distributed, FLOC algorithm is proposed based on the hesitant fuzzy linguistic term set (HFLTS) analysis in order to resolve the CH decision making problems and network lifetime bottlenecks using a dynamic network architecture involving opportunistic clustering. The attributes such as energy transfer based opportunistic routing, energy welfare, relative thermal entropy; expected optimal hops and link quality factor are utilized to form the criteria for Hesitant Fuzzy Linguistic Term Set and make a decision about the contemporary role of the node based on its current state. The effectiveness of FLOC is confirmed after carefully analyzing and evaluating its performance against several existing benchmarks. The simulation results have clearly shown that employing FLOC algorithm results in the improvement of active node ratio, average energy consumption and packet delivery ratio. The possible future work would be to perform the hesitant fuzzy linguistic term set analysis for harvested energy scavenging and transfer capabilities in opportunistic clustering.

**ACKNOWLEDGEMENTS**

The authors would like to thank all the reviewers for their insightful comments and deliberate suggestions for improving the quality of this paper. This work was supported in part by the National Key Research and Development Program of China under Grant 2020YFB1901900, and in part by the National Natural Science Foundation of China (NSFC) under Grant 51776050 and Grant 51536001.

## AUTHORS

**Junaid Anees** received B.S. degree from Institute of Space Technology, Islamabad, Pakistan in 2010 and M.S. degree in Electrical Engineering from COMSATS University Islamabad, Pakistan in 2015. Currently, he is a PhD scholar in School of Energy Science & Engineering at Harbin Institute of Technology, China. He holds Senior Manager Position in Ground Segment Network Operations in Public Sector Organization in Pakistan. His research interest includes Energy harvesting Wireless Sensor Networks, Opportunistic Routing, Smart Grids, and Distributed Computing. He is also interested in Cognitive Radio Sensor Networks and Mobile Networking.

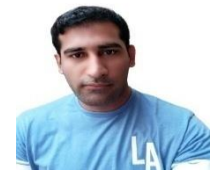

**Prof. & Dr. Hao-Chun Zhang** is currently the Head of the Department of Nuclear Science and Engineering and executive professor of HIT-CORYS Nuclear System Simulation International Joint Research Center (Sino-France). With BE in 1999, ME in 2001 and PhD in 2007 from Harbin Institute of Technology (HIT), Dr. Zhang joined HIT in September 2004. Dr. Zhang has about 200 research publications in peer reviewed journals and conferences, 5 books, and 2 translations of foreign books.

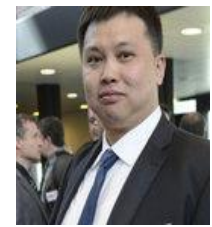